\PassOptionsToPackage{utf8}{inputenc}
\documentclass{bioinfo}
\copyrightyear{2015} \pubyear{2015}

\usepackage[ruled,vlined,linesnumbered]{algorithm2e}
\usepackage{comment}
\SetCommentSty{mycommfont}
\SetKwComment{Comment}{$\triangleright$\ }{}

\newtheorem{definition}{Definition}[section]

\usepackage{url}
\usepackage{makecell}

\access{Advance Access Publication Date: Day Month Year}
\appnotes{Manuscript Category}

\begin{document}
\firstpage{1}

\subtitle{Subject Section}

\title[Graph wavefront algorithm]{Fast sequence to graph alignment using the graph wavefront algorithm}
\author[Zhang \textit{et~al}.]{Haowen Zhang\,$^{\text{\sfb 1}}$, Shiqi Wu\,$^{\text{\sfb 2}}$, Srinivas Aluru\,$^{\text{\sfb 1,3,}*}$ and Heng Li\,$^{\text{\sfb 4,5,}*}$}
\address{$^{\text{\sf 1}}$School of Computational Science and Engineering, Georgia Institute of Technology, USA,\\
$^{\text{\sf 2}}$Department of Mathematics, Nanjing University, China,\\
$^{\text{\sf 3}}$Institute for Data Engineering and Science, Georgia Institute of Technology, USA,\\
$^{\text{\sf 4}}$Department of Data Sciences, Dana-Farber Cancer Institute, USA and\\
$^{\text{\sf 5}}$Department of Biomedical Informatics, Harvard Medical School, USA.
}

\corresp{$^\ast$To whom correspondence should be addressed.}

\history{Received on XXXXX; revised on XXXXX; accepted on XXXXX}

\editor{Associate Editor: XXXXXXX}

\abstract{
\textbf{Motivation:} A pan-genome graph represents a collection of genomes and encodes sequence variations between them. It is a powerful data structure for studying multiple similar genomes. Sequence-to-graph alignment is an essential step for the construction and the analysis of pan-genome graphs. However, existing algorithms incur runtime proportional to the product of sequence length and graph size, making them inefficient for aligning long sequences against large graphs. \\
\textbf{Results:} We propose the graph wavefront alignment algorithm (Gwfa), a new method for aligning a sequence to a sequence graph. Although the worst-case time complexity of Gwfa is the same as the existing algorithms, it is designed to run faster for closely matching sequences, and its runtime in practice often increases only moderately with the edit distance of the optimal alignment. On four real datasets, Gwfa is up to four orders of magnitude faster than other exact sequence-to-graph alignment algorithms. We also propose a graph pruning heuristic on top of Gwfa, which can achieve an additional $\sim$10-fold speedup on large graphs. \\
\textbf{Availability:} Gwfa code is accessible at \url{https://github.com/lh3/gwfa}.\\
\textbf{Contact:} \href{hli@ds.dfci.harvard.edu}{hli@ds.dfci.harvard.edu} and \href{aluru@cc.gatech.edu}{aluru@cc.gatech.edu}\\\
\textbf{Supplementary information:} Supplementary data are available at  \textit{Bioinformatics}
online.} 

\maketitle

\section{Introduction}
Pairwise sequence alignment is an ubiquitous and fundamental problem in bioinformatics. Two sequences of length $N$ can be aligned by dynamic programming (DP)~\citep{needleman1970general} in $O(N^2)$ time. This quadratic runtime is unaffordable for long sequences, motivating the development of diagonal alignment algorithms~\citep{myers1986ano,ukkonen1985algorithms,landau1989fast}. These methods only compute part of the DP matrix within a certain band, thereby reducing the runtime to $O(ND)$ where $D$ is the minimum edit distance between the sequences being aligned. Recently, such methods have been further generalized to work with arbitrary cost linear or affine gap penalty~\citep{xin2017leap,marco2021fast,eizenga2022improving,marco2022optimal}, and named wavefront alignment (WFA) as they progressively compute the partial alignments for increasing alignment scores until the best alignment is found. Compared with the standard DP alignment algorithm, these diagonal alignment methods have shown superior performance in practice when the two sequences to align are similar to each other.

Recent advances in long-read sequencing technologies and assembly methods have enabled the generation of human genome assemblies that meet or even exceed the quality of the reference human genome~\citep{shafin2020nanopore,cheng2021haplotype,nurk2022complete}. As the set of available high-quality human genome assemblies grows, it is important to shift from a single reference genome to a new reference genome model that can support the representation and analysis of the variants from a collection of assembled haplotypes~\citep{wang2022human}. In this transition, pan-genomics has emerged and expanded rapidly as a new research subarea of computational genomics~\citep{computational2018computational}. The idea currently gaining momentum is to replace the single reference genome with a graphical genome model, or a genome graph, which encodes variations among individual genomes in the population~\citep{li2020design,eizenga2020pangenome}. Recent studies have shown that various bioinformatics applications such as read mapping, variants calling, and genotyping can benefit from this transition~\citep{paten2017genome,garrison2018variation,siren2021pangenomics,ebler2022pangenome}.

In the aforementioned applications, aligning sequences to the graph-based structure plays an essential role. This fundamental problem and its variants have been studied by multiple previous works~\citep{lee2002multiple,antipov2016hybridspades,kavya2019sequence,rautiainen2019bit,jain2020complexity}. Besides, this problem has been investigated even earlier in the string literature as approximate pattern matching to hypertext~\citep{manber1992approximate,amir2000pattern,navarro2000improved}. A summary of these results can be found in previous work~\citep{jain2020complexity}. 

Although several sequence to graph alignment algorithms have been proposed, and some of them achieve asymptotically optimal runtime under the Strong Exponential Time Hypothesis~\citep{backurs2015edit}, they are computationally intensive taking $O(|V|+|q||E|)$ time, where $|V|$ and $|E|$ are the numbers of vertices and edges in the graph and $|q|$ is the query length. This motivated the development of GraphAligner~\citep{rautiainen2020graphaligner} and Astarix~\citep{ivanov2020astarix,ivanov2022fast}, which introduced heuristics to accelerate sequence to graph alignment. GraphAligner has implemented a dynamic banded sequence to graph alignment algorithm, which computes cells in the DP matrix with values less than the minimum cost plus a fixed threshold. As the optimal alignment may not be found within the confines of the band, this approach may report suboptimal alignments. Astarix reformulates the sequence to graph alignment as a shortest path problem and uses the A* algorithm to search for it. The A* search algorithm is an extension of Dijkstra's algorithm, and uses a heuristic function to estimate the distance from the vertex currently being explored to the target vertex for pruning the search space. Though the heuristics still guarantee finding the optimal alignment, extra computation cost is needed to process the query or the graph, which can be avoided.

In this paper, we propose the graph wavefront algorithm (Gwfa) for sequence-to-graph alignment to realize benefits analogous to what WFA algorithms achieved for sequence to sequence alignment. Our algorithm runs faster for well-aligned sequences and guarantees alignment optimality in all cases. We present additional heuristics to reduce the alignment runtime and methods to trace the optimal alignment walk in the graph. We construct graphs for biologically important regions rich in polymorphism and empirically demonstrate advantages of Gwfa in terms of runtime and memory usage compared to other graph alignment methods.
\section{Methods}

\subsection{Problem formulation}
Let $\Sigma$ denote an alphabet, and $x$ and $y$ be strings over $\Sigma$. We use $x[i]$ to denote the $i$-{th} character of $x$, $|x|$ to denote its length, and $xy$ to denote the concatenation of $x$ and $y$. Let $x[i,j]$ ($1 \leq i \leq j \leq |x|$) denote $x[i]x[i+1]\ldots x[j]$, the substring of $x$ beginning at the $i$-{th} position and ending at the $j$-{th} position. We assume a constant sized alphabet for DNA sequences, i.e., $\Sigma = \{A,C,G,T\}$.

\begin{definition}[Sequence graph]
A sequence graph $G(V,E,\sigma)$ is a directed graph with vertices $V$ and edges $E$. Function $\sigma:V\rightarrow \Sigma^*$ labels each vertex $v \in V$ with a string $\sigma(v)$ over the alphabet $\Sigma$.
\end{definition}

This definition encompasses commonly used graphs (e.g., de Bruijn graphs, variation graphs) as shown by~\cite{jain2020complexity}. To simplify notation, we may directly use $v$ to denote $\sigma(v)$ and thus $|v|=|\sigma(v)|$ and $v[i]=\sigma(v)[i]$.

Let $w_{j_1,j_n}=v_{j_1}v_{j_2} \ldots v_{j_n}$ ($n \geq 1$) denote a walk that starts from vertex $v_{j_1}$ and ends at vertex $v_{j_n}$ in $G$. It spells the sequence $\sigma(w_{j_1,j_n})=\sigma(v_{j_1})\sigma(v_{j_2}) \ldots \sigma(v_{j_n})$. Let $d(x,y)$ denote the edit distance between strings $x$ and $y$. Then we formulate the problem of optimal sequence to graph alignment as the following.

\begin{definition}[Optimal global sequence to graph alignment]
Given a query sequence $q$, a sequence graph $G(V,E,\sigma)$, a start vertex $v_s\in V$ and an end vertex $v_e \in V$ that is reachable from $v_s$, find a walk $w_{s,e}$ such that $\forall w'_{s,e}$, $d(\sigma (w),q)\leq d(\sigma(w'),q)$.
\end{definition}

This formulation is similar to the global pairwise sequence alignment problem, where the gaps at the beginning or the end of either the query or the target sequence (the walk in the graph) are penalized. Besides, the formulation can be slightly changed to allow gaps at the start or the end of the walk in the sequence graph or perform alignment extension without specifying an end vertex (Supplementary Section 1). These alternative formulations are frequently used to compute the base-level alignments in seed-and-extend methods for read mapping to linear or graphical genomes~\citep{li2009fast,li2018minimap2,rautiainen2020graphaligner,zhang2021fast} when the alignment candidate regions are known as a prior from the seeding step.

\subsection{Sequence to graph alignment algorithm}
\cite{rautiainen2019bit} generalized the standard DP for pairwise sequence alignment to formulate a recurrence for the sequence to graph alignment problem. Though there can be cyclic dependencies in their proposed recurrence, they proved that there is only one unique solution for the optimal alignment cost. Moreover, the recurrence can be solved in $O(|V|+|q||E|)$ time by Navarro's algorithm~\citep{navarro2000improved}. Note that their problem formulations operate on character labeled vertices, without loss of generality. However, directly allowing string labels results in more compact graph representation as the vertices of sequence graphs are usually labeled with strings (long segments of sequences) in practice (e.g., GFA format at \url{https://github.com/GFA-spec/GFA-spec}). Therefore, we first generalize the DP algorithm and propose the following recurrence for sequence alignment to string-labeled sequence graphs.

Let $H_{i,v,j}$ denote the minimum edit distance between a query prefix that ends at its $i$-th position and a walk sequence that ends at $j$-th position of (the string at) vertex $v$, where $0\le i\le|q|$, $v\in V$, and $0\le j\le|v|$. It can be calculated as:
\begin{align}\label{eq:s2gdp}
  \begin{split}
  H_{i, v, j} & =\min
    \begin{cases}
     H_{i - 1, v, j} + 1, & i \geq 1\\ 
     H_{i, v, j-1} + 1, & j \geq 1\\ 
     H_{i-1, v, j-1}+ \Delta_{i, v,j}, & i \geq 1, j \geq 1\\ 
     H_{i, u, |u|}, & j=0, \forall u, (u,v)\in E 
    \end{cases}
  \end{split}
\end{align}
where $\Delta_{i,v,j} = 0$ if $q[i]=v[j]$ or 1 otherwise. When searching for optimal global sequence to graph alignment with start vertex $v_s$ and end vertex $v_e$, the initial condition is $H_{0,v_s,0} = 0$, and the minimum cost is $H_{|q|,v_e,|v_e|}$.




In previous work, \cite{jain2020complexity} have shown that a string labeled sequence graph can be converted to an equivalent character labeled sequence graph by splitting the vertex with a string label into a chain of character labeled vertices to compute sequence-to-graph alignment. Similarly, we argue that a character labeled graph can be converted to an equivalent string labeled graph by compacting each chain of character labeled vertices into a single vertex with string label that is the concatenation of the individual character labels, while the in-edges of the first vertex in the chain and the out-edges of the last vertex are reflected in how the new vertex is connected. Given these transformations, the proposed recurrence for sequence alignment to a string labeled graph is essentially equivalent to the recurrence for sequence alignment to a character labeled graph. Therefore, the algorithm proposed by~\cite{navarro2000improved} can also be adapted to solve the recurrence in Equation~\ref{eq:s2gdp}, which we show in Supplementary Section 2.

\subsection{Diagonal recurrence}
We use $(i,v,j)$ to denote a DP-cell in $H$, where $0\le i\le|q|$, $v\in V$, and $0\le j\le|v|$. Note that the three-dimensional DP matrix $H$ can be regarded as a set of $|V|$ two-dimensional DP matrices $\{H_v | v\in V\}$ along the vertex dimension. And $H_v$ contains DP-cells $\{(i,v,j)|0\le i\le|q|,0\le j\le|v|\}$. Let $(v,k)$ denote a diagonal in $H_v$. A DP cell $(i,u,j)$ is on diagonal $(v,k)$ if and only if $u=v$ and $k=i-j$. Therefore, DP cell $(k+j,v,j)$ is always on diagonal $(v,k)$. Define
$$
\tilde{H}_{d,v,k}=\left\{\begin{array}{ll}
\max\{j\;|\;H_{k+j,v,j}=d\}, & \mbox{if $
\exists j$ \;s.t.\; $H_{k+j,v,j}=d$}\\
\infty , & \mbox{otherwise}
\end{array}
\right.
$$
which is the furthest offset on diagonal $(v,k)$ among DP cells with edit distance $d$. Importantly, if $\tilde{H}_{d,v,k}=j$, it is always true that $H_{k+j,v,j}=d$, but conversely, if $H_{k+j,v,j}=d$, we only have $j\le\tilde{H}_{d,v,k}$ because multiple cells on diagonal $(v,k)$ may have the same distance $d$.

Following the WFA formulation, we calculate $\tilde{H}_{d,v,k}$ by rewriting Equation~\ref{eq:s2gdp} into its diagonal recurrence as Equation~\ref{eq:s2gdiagdp}:
\begin{align}\label{eq:s2gdiagdp}
\begin{split}
  \tilde{J}_{d, v, k} = & \max
    \begin{cases}
     \tilde{H}_{d-1, v, k - 1}\\
     \tilde{H}_{d-1, v, k + 1} + 1\\
     \tilde{H}_{d-1, v, k} + 1 \\
     0, \hspace{0.6cm}\exists u, (u,v)\in E, \tilde{H}_{d, u, k - |u|} = |u|\\
    \end{cases}\\
  \tilde{H}_{d, v, k} = &
     j+{\rm LCP}\left(q[k+j+1, |q|], v[j+1,|v|]\right), 
     j=\tilde{J}_{d, v, k} \\
\end{split}
\end{align}
Here ${\rm LCP}(x,y)$ gives the length of the longest common prefix between two strings $x$ and $y$. The initial condition is $\tilde{J}_{0,v_s,0} = 0$ for global sequence to graph alignment problem, and we aim to find the minimum edit distance $\hat{d}$ such that $\tilde{H}_{\hat{d},v_e,|q|-|v_e|}=|v_e|$, i.e. $H_{|q|,v_e,|v_e|}=\hat{d}$.

Essentially, the first three cases for $\tilde{J}_{d,v,k}$ in Equation~\ref{eq:s2gdiagdp} are equivalent to running the wavefront algorithm within a vertex. The last case corresponds to the last case in Equation~\ref{eq:s2gdp}. To see the connection, we note that $\tilde{H}_{d,u,i-|u|}=|u|$ implies $H_{i,u,|u|}=d$ and $\tilde{H}_{d,v,i}=0$ implies $H_{i,v,0}=d$. When $u$ is a predecessor of $v$, we carry the computation at the last position of $u$ onto the $0^{th}$ position of $v$. The ${\rm LCP}(\cdot,\cdot)$ term extends along exact matches such that we can find the furthest cell in accordance with the definition of $\tilde{H}_{d,v,k}$.

To explain the implementation of Equation~\ref{eq:s2gdiagdp} in the next section, we further define $d$-wave
$\mathcal{W}_d=\{(i,v,j)|H_{i,v,j}=d\}$,
which consists of DP cells with the value $d$. Then $(k+\tilde{H}_{d,v,k},v,\tilde{H}_{d,v,k})\in\mathcal{W}_d$ is at the front of the $d$-wave along diagonal $(v,k)$, and it is called a graph wavefront. The $d$-wavefront is
$
\mathcal{F}_d=\{(k+j,v,j)\;|\;j=\tilde{H}_{d,v,k}<\infty,-|v|\le k\le|q|,v\in V\}
$.



\subsection{Graph wavefront algorithm}
Our method for computing optimal global sequence to graph alignment using the diagonal formulation is presented in Algorithm~\ref{algo:gwfed}. We assume there is a walk from the start vertex $v_s$ to the end vertex $v_e$. The algorithm progressively increases the edit distance $d$. In each iteration, it finds the $d$-wavefront $\mathcal{F}_d$ with Algorithm~\ref{algo:gwfextend}. It then uses Algorithm~\ref{algo:gwfexpand} to collect cells in the ($d$+1)-wave that are adjacent to $\mathcal{F}_d$. This process is repeated until the whole query is aligned to a walk with the last character of $v_e$ as the end (line 8 in Algorithm~\ref{algo:gwfed}).


\begin{algorithm}[!hb]
\DontPrintSemicolon
\footnotesize
\KwIn{Query sequence $q$, sequence graph $G=(V,E,\sigma)$, start vertex $v_s\in V$ and end vertex $v_e \in V$.}
\BlankLine
\textbf{function} {\sc GwfEditDist}$(q, G, v_s, v_e)$
\Begin {
	$k_{e} \gets |q|-|v_e|$\;
	$\tilde{H}_{v_s,0} \gets 0$\;
	$Q \gets [(v_s, 0)]$\;
	$d\gets0$\;
	\While{{\bf true}} {
		{\sc GwfExtend}$(q,G,Q,\tilde{H})$\;
		\If {$\tilde{H}_{v_e, k_{e}} = |v_e|$ } {
			{\bf return} $d$\;
		}
		$d\gets d+1$\;
		{\sc GwfExpand}$(q,G,Q,\tilde{H})$\;
	}
}
\caption{Graph wavefront algorithm to find the optimal global sequence to graph alignment}\label{algo:gwfed}
\end{algorithm}


Specifically, in the iteration for distance $d$, the part of $d$-wave adjacent to $\mathcal{F}_{d-1}$ is first extended along each diagonal $(v,k)$ through exact matches (lines 5--8 of Algorithm~\ref{algo:gwfextend}). If the end of vertex $v$ is reached, lines 11--14 in Algorithm~\ref{algo:gwfextend} traverse $v$'s neighbors and prepare potential extensions in them. This extension step finds the $d$-wavefront $\mathcal{F}_d$. Then lines 8 and 9 in Algorithm~\ref{algo:gwfed} check $\mathcal{F}_d$ and return the optimal alignment cost when $(|q|,v_e,|v_e|)\in \mathcal{F}_d$. If $(|q|,v_e,|v_e|)\notin \mathcal{F}_d$, Algorithm~\ref{algo:gwfexpand} next finds the set of DP cells adjacent to $\mathcal{F}_d$. These cells are part of the ($d$+1)-wave. They will be extended in the next iteration. We keep track of the graph wavefront in the alignment process by a set and a queue. The set stores the offsets (i.e., $\tilde{H}$) and allows the access of any offset with its diagonal in constant time. This graph wavefront set can be implemented with an array of which the size is the number of diagonals and a special sign to mark the existence of an element in the set. The queue keeps track of the diagonals on which the graph wavefront can be further updated or used to update the graph wavefront on other diagonals. Thus, only the graph wavefront on the diagonals in the queue instead of all the diagonals are processed in each iteration.


\begin{algorithm}[!hb]
\DontPrintSemicolon
\footnotesize
\KwIn{Query sequence $q$, sequence graph $G=(V,E,\sigma)$, queue $Q$ to keep track of diagonals and graph wavefront set $\tilde{H}$.}
\BlankLine
\textbf{function} {\sc GwfExtend}$(q,G,Q,\tilde{H})$
\Begin {
    $Q' \gets [\ ]$\;
	\While {$Q$ is not empty} {
	    $(v,k) \gets Q.{\rm pop}()$\;
	    $j \gets \tilde{H}_{v,k}$\;
		$i \gets k+j$\;
		$l \gets {\rm LCP}(q[i+1, |q|], v[j+1, |v|])$\;
		$\tilde{H}_{v,k} \gets j+l$\;
		$Q'.{\rm push}(v,k)$\;
		\If {$\tilde{H}_{v,k} = |v|$} {
			\For {$(v,u) \in E$} {
			    \If {$\tilde{H}_{u,k+|v|} \notin \tilde{H}$} {
			        $\tilde{H}_{u,k+|v|} \gets 0$\;
			        $Q.{\rm push}(u,k+|v|)$\;
			    }
			}
		}
	}
	$Q\gets Q'$\;
}
\caption{Graph wavefront extension algorithm}\label{algo:gwfextend}
\end{algorithm}

\begin{algorithm}[!hb]
\DontPrintSemicolon
\footnotesize
\KwIn{Query sequence $q$, sequence graph $G=(V,E,\sigma)$, queue $Q$ to keep track of diagonals and graph wavefront set $\tilde{H}$.}
\BlankLine
\textbf{function} {\sc GwfExpand}$(q,G,Q,\tilde{H})$
\Begin {
    $\tilde{H}' \gets \emptyset$\;
    $Q' \gets [\ ]$\;
	\While {$Q$ is not empty} {
	    $(v,k) \gets Q.{\rm pop}()$\;
		$i \gets k+\tilde{H}_{v,k}$\;
		\If {$i < |q|$} {
		    \uIf{$\tilde{H}'_{v,k+1} \notin \tilde{H}'$} {
		        \If{$\tilde{H}_{v,k+1} \notin \tilde{H}$ {\bf or} $\tilde{H}_{v,k + 1} < \tilde{H}_{v,k}$} {
		        	$\tilde{H}'_{v,k+1} \gets \tilde{H}_{v,k}$\;
		            $Q'.{\rm push}(v,k+1)$\;
		        }
		    } \Else {
		    	$\tilde{H}'_{v,k+1}\gets \max\{\tilde{H}'_{v,k+1},\tilde{H}_{v,k}\}$\;
		    }
	    }
	    \If {$\tilde{H}_{v,k}< |v|$} {
	        \uIf{$\tilde{H}'_{v,k-1} \notin \tilde{H}'$} {
	        	\If{$\tilde{H}_{v,k-1} \notin \tilde{H}$ {\bf or} $\tilde{H}_{v,k - 1} < \tilde{H}_{v,k} + 1$} {
	        	    $\tilde{H}'_{v,k-1} \gets \tilde{H}_{v,k} + 1$\;
		            $Q'.{\rm push}(v,k-1)$\;
		        }
	        } \Else {
	            $\tilde{H}'_{v,k-1}\gets \max\{\tilde{H}'_{v,k-1}, \tilde{H}_{v,k} + 1\}$\;
	        }
		}
		\If {$i < |q|$ {\bf and} $\tilde{H}_{v,k}< |v|$} {
		    \uIf{$\tilde{H}'_{v,k} \notin \tilde{H}'$} {
		    	$\tilde{H}'_{v,k} \gets \tilde{H}_{v,k} + 1$\;
		        $Q'.{\rm push}(v,k)$\;
		    } \Else {
		        $\tilde{H}'_{v,k}\gets \max\{\tilde{H}'_{v,k}, \tilde{H}_{v,k} + 1\}$\;
		    }
		}
	}
    \For{$(v,k)\in Q'$}{
        $\tilde{H}_{v,k} \gets \tilde{H}'_{v,k}$\;
    }
    $Q\gets Q'$\;
}
\caption{Graph wavefront expansion algorithm}\label{algo:gwfexpand}
\end{algorithm}

To analyze the runtime of our proposed algorithm, we compute the time spent on updating the graph wavefront and exploring neighbors. After one round of extension and expansion, the graph wavefront on the diagonals in the queue advances at least one DP-cell forward if it has not reached the end. Once the graph wavefront has already reached the last cell on a diagonal, it cannot be updated using the graph wavefront on other diagonals. Thus the number of updates on each diagonal is bounded by the length of the diagonal, which means the total number of updates is bounded by the total length of the diagonals as $O(|q|\sum_{v\in V}|v|)$. 

Next, we compute the time spent on neighbor exploration (lines 10-14 in Algorithm~\ref{algo:gwfextend}). As mentioned above, on each diagonal, the graph wavefront only reaches the end of the diagonal once. So the neighbors of each vertex are explored only once in the extension step when the graph wavefront reaches the end of the diagonal of the vertex. We divide all diagonals into $|q|$ sets, $K_1,\ldots, K_{|q|}$, where $K_i$ contains diagonals $\{(v, i-|v|)|v\in V\}$. As $K_i$ has one diagonal for each vertex, the cost to explore the neighbors of vertices that have corresponding diagonals in $K_i$ is $O(|E|)$. Thus the total time spent on neighbor exploration is $O(|q||E|)$. Therefore, the overall runtime is $O(|q|(\sum_{v\in V}|v| + |E|))$. Since exact and approximate sequence matching to graphs are equally hard~\citep{equi2019complexity}, our proposed algorithm is asymptotically optimal under SETH~\citep{backurs2015edit}. The memory usage is $O(|q||V| + \sum_{v\in V}|v|)$ since both the size of the queue to keep track of the diagonals and the graph wavefront set size are bounded by the total number of diagonals.

Note that Algorithms~\ref{algo:gwfed} and \ref{algo:gwfextend} can also be slightly adapted to solve other sequence to graph alignment problem variants, which we show in Supplementary Section 1.

\subsection{Graph wavefront pruning}
The size of the $d$-wave increases with edit distance $d$. While some promising components on the $d$-wavefront $\mathcal{F}_d$ are advancing towards the solution, many other components on $\mathcal{F}_d$ significantly fall behind those promising ones. Based on this observation, we propose a graph wavefront pruning heuristic (Algorithm~\ref{algo:gwfprune}) to eliminate those unpromising graph wavefront components, thereby accelerating the alignment process. This algorithm is similar to the wavefront reduction heuristic used by WFA, and it may miss the optimal alignment.

The pruning algorithm first finds the maximum sum of aligned query length and aligned graph walk length $a_{\rm max}$ from all the DP-cells in the graph wavefront. Then each graph wavefront is rechecked to see whether it is left behind too much. If its sum of aligned query length and aligned graph walk length is too short compared with $a_{\rm max}$, the graph wavefront will be dropped. Since promising graph wavefronts usually start to appear after several iterations, the check is only performed when the offsets of the graph wavefront are already large enough to avoid overhead. The pruning can be performed before the graph wavefront expansion step (between line 10 and 11 in Algorithm~\ref{algo:gwfed}), which would reduce the size of the graph wavefront set to expand.

\begin{algorithm}[!hb]
\DontPrintSemicolon
\footnotesize
\KwIn{Max allowed difference $a_{\rm diff}$ from the max sum of aligned query length and aligned graph walk length, queue $Q$ to keep track of diagonals, graph wavefront set $\tilde{H}$, and aligned graph walk length set $\tilde{W}$.}
\BlankLine
\textbf{function} {\sc GwfPrune}$(a_{\rm diff}, Q, \tilde{H}, \tilde{W})$
\Begin {
	$a_{\rm max} \gets 0$\;
	\For {$(v,k)\in Q$} {
		$i \gets k+\tilde{H}_{v,k}$\;
		$a \gets i + \tilde{W}_{v,k}$\;
		$a_{\rm max} \gets \max(a_{\rm max},a)$\;
	}
	\If {$a_{\rm max} > a_{\rm diff}$} {
	$\tilde{H}' \gets \emptyset$\;
	$Q'\gets [\ ]$\;
	\While {$Q$ is not empty} {
	    $(v,k) \gets Q.{\rm pop()}$\;
		$i\gets k+\tilde{H}_{v,k}$\;
		$a \gets i + \tilde{W}_{v,k}$\;
		\If {$a_{\rm max} - a < a_{\rm diff}$} {
		    $\tilde{H}'_{v,k} \gets \tilde{H}_{v,k}$\;
		    $Q'.{\rm push}(v,k)$\;
		}
	}
	\For{$(v,k)\in Q'$}{
        $\tilde{H}_{v,k} \gets \tilde{H}'_{v,k}$\;
    }
	$Q \gets Q'$\;
	}
}
\caption{Graph wavefront pruning algorithm}\label{algo:gwfprune}
\end{algorithm}

\subsection{Graph alignment walk traceback}
We further present Algorithm~\ref{algo:gwfpushtrace} and~\ref{algo:gwftraceback} to compute the optimal alignment walk in the graph. We slightly modified Algorithms~\ref{algo:gwfed} and~\ref{algo:gwfextend} to save traceback information during the graph extension process (Supplementary Section 3). Then the alignment walk can be computed with the traceback information. Algorithm~\ref{algo:gwfpushtrace} is used to record the traceback information when exploring the neighbors of the vertices. It saves information of the previous vertex that leads to the extension to the current vertex. After the minimum edit distance is found, we can start from the last vertex in the alignment walk and iteratively trace vertices recorded in the extension process (Algorithm~\ref{algo:gwftraceback}).

\begin{algorithm}[!hb]
\DontPrintSemicolon
\footnotesize
\KwIn{Traceback information array $T$, the vertex $v$ to record traceback information, the previous traceback information index $t$ in $T$ and a hash map $M$ to avoid duplicate traceback information.}
\BlankLine
\textbf{function} {\sc GwfPushTrace}$(v, t, T, M)$
\Begin {
    \If{$M_{v,t} \notin M$}{
        $M_{v,t}\gets T.{\rm size}()$\;
        $T.{\rm push}(v,t)$\;
    }
	{\bf return} $M_{v,t}$\;
}
\caption{Method to add a piece of traceback information during the alignment}\label{algo:gwfpushtrace}
\end{algorithm}

\begin{algorithm}[!hb]
\DontPrintSemicolon
\footnotesize
\KwIn{Traceback information array $T$ and the walk end index $t_e$ in $T$.}
\BlankLine
\textbf{function} {\sc GwfTraceback}$(t_e, T)$
\Begin {
    $W \gets [\ ]$\;
    $t \gets t_e$\;
	\While {$t \geq 0$ {\bf and} $T[t].v \geq 0$} {
	    $W.{\rm push}(T[t].v)$\;
	    $t \gets T[t].t$\;
	}
	Reverse $W$\;
	{\bf return} $W$\;
}
\caption{Graph alignment walk traceback algorithm}\label{algo:gwftraceback}
\end{algorithm}
\section{Results}
We implemented our proposed method and termed it Gwfa, made available at \url{https://github.com/lh3/gwfa}. In the following sections, we establish benchmark datasets and demonstrate the advantages of Gwfa empirically compared with other methods.

\subsection{Experimental setup}
\subsubsection{Benchmark data sets}
To evaluate the performance of Gwfa, we obtained four sequence graphs around complement component 4 (C4), leukocyte receptor complex (LRC), and major histocompatibility complex (MHC) loci in the human genome (Table~\ref{table:graphs}). 
\begin{table}[htbp]
\processtable{List of benchmark sequence graphs.}{\label{table:graphs}
\begin{tabular}{lllrrr}
\toprule
\thead{Graph} &{Type} & \thead{Region} & \thead{Total segment\\ length (bp)} & \thead{\# vertices} & \thead{\# edges}\\
\midrule
G1 & Cyclic & C4 &	42,036 &	1,531 &	2,073\\
G2 & Cyclic & LRC &	1,294,511 &	48,097 &	67,008\\
G3 & Cyclic & MHC &	5,951,398 &	232,508 &	320,009\\
G4 & Acyclic & MHC &	5,476,947 &	1,144 &	1,608\\
\midrule
\end{tabular}}{}
\end{table}
These three loci play crucial biological roles and are enriched with polymorphisms. The numerous variations at these loci increases the complexity of the graphs which poses a great challenge to alignment. As a result, these loci are often used to benchmark graphical genome tools~\citep{jain2019accelerating,jain2019validating,li2020design,guarracino2021odgi}. Specifically, MHC is one of the most critical regions for infection and autoimmunity in the human genome and is crucial in adaptive and innate immune responses~\citep{horton2004gene}. MHC haplotypes are highly polymorphic, i.e., various alleles are present across individuals in the population. C4 genes are located in MHC and encode proteins involved in the complement system. The LRC locus contains many genetic variations and comprises genes that can regulate immune responses~\citep{barrow2008extended}.

To construct G1, G2 and G3, we used ODGI~\citep{guarracino2021odgi} to extract the subgraphs around the C4 (chr6:31,972,057--32,055,418 on GRCh38), LRC (chr19:54,528,888--55,595,686) and MHC (chr6:29,000,000--34,000,000) loci from the PGGB human pangenome graphs released by the Human Pangenome Reference Consortium (HPRC). The PGGB graph was built from GRCh38, CHM13~\citep{nurk2022complete}, and the phased contig assemblies of 44 diploid individuals. These altogether encode variations from 90 human haplotypes.
To construct G4, we used gfatools~(\url{https://github.com/lh3/gfatools}) to extract the MHC region from the minigraph graph~\citep{li2020design} built from the same set of assemblies. G4 contains fewer vertices because the minigraph graph only encodes $\ge$50 bp structural variations, while the PGGB graph additionally encodes SNPs and short INDELs in all input samples. The details and command lines to build these sequence graphs can be found in Supplementary Section 4.


As HG002, HG005, and NA19240 diploid assemblies were excluded when building the human pangenome graphs, these six haplotypes provide ideal query sequences for the alignment evaluation. We ran ``minimap2 -x asm20'' to align the C4, LRC, and MHC regions of GRCh38 to HG002, HG005, and NA19240 haplotypes, and retrieved the corresponding regions from the six haplotypes. We excluded LRC on the HG002 maternal haplotype because it was not fully assembled.

\subsubsection{Hardware and software}
We ran all experiments on a compute node with dual Intel Xeon Gold 6226 CPU (2.70 GHz) processors equipped with 128 GB main memory. The time and memory usage of each run was measured.

We ran Gwfa both in its exact mode and approximate mode with the pruning heuristic, which is controlled by the $a_{\rm diff}$ parameter. We set this parameter to 20,000 for the G1--G3 dataset and 30,000 for the G4 dataset. Setting $a_{\rm diff}$ to 20,000 for the G4 dataset would miss the optimal alignment. We also evaluated Gwfa with traceback to output the walk corresponding to the optimal alignment identified.

Sequence-to-graph alignment can be formulated as a shortest path problem~\citep{jain2020complexity}. In this work, we implemented this idea on top of Dijkstra's algorithm. We also generalized Navarro's algorithm for string-labeled graphs and provided an efficient implementation (Supplementary Algorithm 3). Both implementations are available at \url{https://github.com/haowenz/sgat}. They serve as a baseline and help to verify the correctness of our graph wavefront algorithm implementation.

For further comparative evaluation, we considered GraphAligner \citep{rautiainen2020graphaligner} and Astarix~\citep{ivanov2020astarix}, two recently developed tools for accelerating sequence to graph alignment. Like Gwfa, both of these can also work with cyclic sequence graphs. GraphAligner implements the bit-parallel algorithm~\citep{rautiainen2019bit}, which guarantees to finding an optimal alignment, as well as a seed-and-extend heuristic method, which is fast in practice. We included both in our evaluation (see Supplementary Section 5 for more details about running the tools). Astarix could in theory find minimum edit distance with option `align-optimal -G 1'. However, Astarix either reported errors or could not finish any of the experiments and was thus excluded.

To investigate the performance gap between aligning a query to a target sequence and a sequence graph, we also ran Edlib~\citep{vsovsic2017edlib} and WFA2~\citep{marco2021fast} to align the C4 regions of HG002, HG005, and NA19240 to the GRCh38 C4 region. The parameters for running these tools are listed in Supplementary Section 6. We compared these two tools with the sequence to graph alignment tools mentioned above on aligning queries to a linear sequence graph built from the GRCh38 C4 region (Supplementary Section 7).

\subsection{Runtime comparison}
Table~\ref{table:graphruntime} shows the runtime of each method. When aligning queries to G1, Gwfa and Gwfa-pruning were several orders of magnitude faster than all other sequence to graph alignment algorithms. This is expected since G1 is built from 90 human haplotypes and encodes all their variants, which leads to relatively small edit distances for all the alignments. Both Gwfa and Gwfa-pruning were able to skip the exploration of many DP-cells, which made them significantly faster than other methods. Due to the same reason, our Dijkstra's algorithm implementation was faster than other methods except for Gwfa and Gwfa-pruning on aligning queries to G1. In addition, all the methods were able to report optimal alignments between the queries and sequence graph G1.

\begin{table*}[htbp]
\processtable{Runtimes (in seconds) of the methods to align queries to real sequence graphs.}{\label{table:graphruntime}
\begin{tabular*}{\textwidth}{@{\extracolsep{\fill}}llr|rrrrrr}
\toprule
Graph & Haplotype & \makecell{Edit\\distance} &Gwfa &\makecell{Gwfa-pruning} & \makecell{Dijkstra's\\algorithm} & \makecell{Navarro's\\algorithm} & \makecell{GraphAligner\\bitvector} & \makecell{GraphAligner\\heuristic}\\
\hline
G1  & HG002.1 & 22   & \textbf{<0.01}   &   \textbf{<0.01} &   0.14   &   17.40   &   4.23 &   1.31\\
    & HG002.2 & 22   & \textbf{<0.01}   &   \textbf{<0.01} &   0.14   &   17.40   &   4.24 &   0.18\\
    & HG005.1 & 21   & \textbf{<0.01}   &   \textbf{<0.01} &   0.10   &   17.40   &   4.24 &   0.18\\
    & HG005.2 & 19   & \textbf{<0.01}   &   \textbf{<0.01} &   0.07   &   17.40   &   4.24 &   0.19\\
    & NA19240.1 & 19  & \textbf{<0.01}   &   \textbf{<0.01} &   0.07   &   15.90   &   3.87 &   0.16\\
    & NA19240.2 & 20   & \textbf{<0.01}   &   \textbf{<0.01} &   0.06   &   15.90   &   3.85 &   0.16\\
\hline
G2 & HG002.1  & 65   & 1.20  &   \textbf{0.13}    &   760.13   &   10,285.00   &   1,186.55   &   14.12\\
    & HG005.1 & 81   & 1.87  &   \textbf{0.15}    &   1189.67   &   10,146.00   &   1,161.97   &   12.99\\
    & HG005.2  & 174   & 5.42  &   \textbf{0.24}    &   3340.85   &   10,045.00   &   1,140.04   &   5.99\\
    & NA19240.1 & 452   & 13.91  &   \textbf{0.62}    &   9,944.00   &   10,184.00   &   1,134.08   &   $^*$5.58\\
    & NA19240.2 & 469   & 10.73  &   \textbf{0.51}    &   6,476.00    &   9,702.00   &   1,110.03   &   9.80\\
\hline
G3 & HG002.1 & 331   & 45.52 &   \textbf{5.40}   &   -   &   -   &   19,389.98   &   24.25\\
    & HG002.2 & 268  & 16.29 &   \textbf{1.76}   &   -   &   -   &   17,594.77   &   29.24\\
    & HG005.1  & 300   & 23.78 &   \textbf{0.91}   &   -   &   -   &   17,754.32   &   26.48\\
    & HG005.2 & 298   & 20.90 &   \textbf{0.69}   &   -   &   -   &   19,229.14   &   22.31\\
    & NA19240.1 & 743   & 68.84 &   \textbf{2.07}   &   -   &   -   &   17,679.07   &   $^*$32.06\\
    & NA19240.2 & 885   & 132.38 &   \textbf{2.06}   &   -   &   -   &  18,215.37   &   30.39\\
\hline
G4 & HG002.1 & 43,280   & 868.94 &   \textbf{7.70}   &   -   &   -   &   15,008.00   &   $^*$31.35\\
    & HG002.2 & 34,790  & 453.07 &   \textbf{4.39}   &   -   &   -   &   14,733.00   &   $^*$29.42\\
    & HG005.1  & 42,124   & 849.21 &   \textbf{6.97}   &   -   &   -   &   15,059.50   &   $^*$315.57\\
    & HG005.2 & 41,271   & 824.15 &   \textbf{6.75}   &   -   &   -   &   15,061.50   &   $^*$28.68\\
    & NA19240.1 & 42,677   & 709.35 &   \textbf{7.43}   &   -   &   -   &   15,017.50   &   $^*$29.63\\
    & NA19240.2 & 31,564   & 346.96 &   \textbf{2.90}   &   -   &   -   &  15,109.50   &   $^*$31.39\\
\midrule
\end{tabular*}}{Note: Dijkstra's algorithm and generalized Navarro's algorithm cannot finish in 24 hours when mapping queries to the MHC graphs. The LRC region of HG002.2 was not fully assembled. The GraphAligner heuristic would fragment the alignment into multiple pieces when its runtime is marked with $^*$.}
\end{table*}

On the G2 dataset, Gwfa-pruning was 9 times to 5 orders of magnitude faster than other methods. Gwfa in its exact mode was even faster than the GraphAligner heuristic on the three query sequences with relatively small edit distances. While the GraphAligner heuristic was faster on the other two queries, it fragmented the alignment of the NA19240.1 haplotype and failed to report the expected alignment.

On the much larger G3 dataset, both Dijkstra's and Navarro's algorithms could not finish in 24 hours. Gwfa-pruning was the fastest. Gwfa in its exact mode generally has comparable performance to the GraphAligner in its heuristic mode. GraphAligner heuristic again fragmented the alignment of the NA19240.1 haplotype.

As G4 only encodes large variants, the edit distances of alignments on G4 are much larger than those on other datasets. Thus, the speed of Gwfa decreased. Although the GraphAligner heuristic was faster than Gwfa, it failed to report complete alignments for all query sequences. The Gwfa pruning heuristic was the fastest and could report the optimal distances.

When the optimal alignment walk was traced, there was only a mild increase in runtime for Gwfa or Gwfa-pruning. When aligning queries to G1, the runtime with alignment walk traceback was still less than 0.01s. On average, the runtime increased by 13.6\% and 3.8\% for Gwfa and Gwfa-pruning to trace the optimal alignment walk on G2, and 14.5\% and 7.8\% on G3 respectively. For alignment traceback on G4, the runtime increased less than 2\% for Gwfa with or without pruning mainly because G4 has fewer vertices and edges than other graphs.

\subsection{Memory usage}
Table~\ref{table:graphmemory} shows the memory usage of the tools. When aligning queries to sequence graphs, Gwfa-pruning used 2-83 times less memory than other methods. Besides, compared with other methods, Gwfa without pruning also used comparable or less memory. The alignment walk traceback process of Gwfa and Gwfa-pruning on G1 needs an extra 6.8\% memory on average. The average memory usage increased by 49.1\% and 15.8\% for Gwfa and Gwfa-pruning to trace back the alignment walk on G2, and 75.2\% and 18.9\% on G3, respectively, as the problem size is larger. Nevertheless, Gwfa-pruning still used the least memory to align any queries to G2 or G3 and trace the alignment. For alignments on G4, the memory usage of Gwfa with or without pruning only increased by less than 3\% on average. This is because G4 has fewer vertices than the other graphs, which results in a smaller traceback stack size. The results consistently point to small memory footprint needed by our proposed methods.
\begin{table*}[htbp]
\processtable{Memory usage (MB) of the methods to align queries to real sequence graphs.}{\label{table:graphmemory}
\begin{tabular*}{\textwidth}{@{\extracolsep{\fill}}llr|rrrrrr}
\toprule
Graph & Haplotype & \makecell{Edit\\distance}& Gwfa &\makecell{Gwfa-pruning} & \makecell{Dijkstra's\\algorithm} & \makecell{Navarro's\\algorithm} & \makecell{GraphAligner\\bitvector} & \makecell{GraphAligner\\heuristic}\\
\hline
G1  & HG002.1  & 22   & 3.3   &   \textbf{2.7} &   191.1   &   165.4   &   7.0 &   53.6\\
    & HG002.2  & 22   & 3.5   &   \textbf{2.5} &   190.8   &   165.4   &   7.0 &   51.1\\
    & HG005.1  & 21   & 3.5   &   \textbf{2.5} &   186.4   &   165.4   &   7.0 &   51.2\\
    & HG005.2 & 19   & \textbf{2.2}   &   \textbf{2.2} &   184.1   &   165.4   &   7.0 &   51.5\\
    & NA19240.1 & 19   & \textbf{2.2}   &   \textbf{2.2} &   184.1   &   165.4   &   7.0 &   50.2\\
    & NA19240.2 & 20   & 2.7   &   \textbf{2.5} &   184.1   &   165.4   &   7.0 &   48.0\\
\hline
G2 & HG002.1  & 65   & 211.4 &   \textbf{37.4}    &   2,364.4   &   213.3   &   164.9   &   476.0\\
    & HG005.1  & 81   & 214.9 &   \textbf{37.4}    &   3,463.8   &   213.3   &   164.6   &   459.3\\
    & HG005.2  & 174   & 212.4 &   \textbf{37.4}    &   8,691.5   &   213.3   &   164.6   &   503.5\\
    & NA19240.1 & 452   & 230.6  &   \textbf{42.9}    &   23,995.3   &   213.3   &   164.7   &   $^*$455.8\\
    & NA19240.2 & 469   & 244.6  &   \textbf{37.4}    &   19,185.0   &   213.3   &   164.5   &   446.3\\
\hline
G3 & HG002.1   & 331   & 975.7 &   \textbf{442.9}   &   -   &   -   &   769.4   &   1251.9\\
    & HG002.2  & 268   & 824.9 &   \textbf{224.8}   &   -   &   -   &   769.0   &   1,388.5\\
    & HG005.1  & 300   & 912.6 &   \textbf{178.1}   &   -   &   -   &   769.2   &   1,587.6\\
    & HG005.2 & 298   & 946.7 &   \textbf{158.0}   &   -   &   -   &   769.2   &   1,232.8\\
    & NA19240.1 & 743   & 960.0 &   \textbf{235.6}   &   -   &   -   &   769.2   &   $^*$1,352.6\\
    & NA19240.2 & 885   & 934.6 &   \textbf{220.8}   &   -   &   -   &   769.4   &   1,566.3\\
\hline
G4 & HG002.1   & 43,280  & 439.1 &   \textbf{32.8}   &   -   &   -   &   889.6   &   $^*$1,122.8\\
    & HG002.2  & 34,790   & 400.0 &   \textbf{27.0}   &   -   &   -   &   889.2   &   $^*$1,041.0\\
    & HG005.1  & 42,124   & 421.2 &   \textbf{27.1}   &   -   &   -   &   889.4   &   $^*$1,063.1\\
    & HG005.2 & 41,271   & 441.2 &   \textbf{32.5}   &   -   &   -   &   889.4   &   $^*$971.7\\
    & NA19240.1 & 42,677   & 425.7 &   \textbf{37.0}   &   -   &   -   &   889.4   &   $^*$1,012.6\\
    & NA19240.2 & 31,564   & 397.8 &   \textbf{36.8}   &   -   &   -   &   889.7   &   $^*$1,113.4\\
\midrule
\end{tabular*}}{Note: Dijkstra's algorithm and generalized Navarro's algorithm cannot finish in 24 hours when mapping queries to the MHC graphs. The LRC region of HG002.2 was not fully assembled. The GraphAligner heuristic would fragment the alignment into multiple pieces when its runtime is marked with $^*$.}
\end{table*}
\section{Discussion}
Sequence-to-graph alignment is the foundation of various pan-genomics applications. However, it is computationally expensive when the sequence is long or the graph is large, as the worst-case runtime is the product of these. Previous algorithms could not fully leverage the similarity information between the sequence and its optimal alignment walk in the graph to reduce runtime while preserving optimality. In this work we proposed Gwfa, a novel sequence-to-graph alignment algorithm, and a heuristic to accelerate the alignment further. We demonstrated empirically the superior performance of Gwfa over other sequence-to-graph alignment algorithms on various input queries and graphs. Although the Gwfa algorithm described here does not directly generate base alignment, we can align the query to the traceback walk with a sequence-to-sequence alignment algorithm to produce such. As sequence alignment is usually many-fold faster than graph alignment, the impact of the additional step on total runtime is negligible.

\section*{Funding}
This work has been supported in part by the National Institutes of Health under U01HG010961 and R01HG01004, and the National Science Foundation under CCF-1816027. 


\bibliographystyle{natbib}

\bibliography{document}

\end{document}